\documentclass[12pt]{article}
\usepackage{amsmath,amsfonts}
\usepackage{algorithmic}
\usepackage{algorithm}
\usepackage{array}
\usepackage[caption=false,font=normalsize,labelfont=sf,textfont=sf]{subfig}
\usepackage{textcomp}
\usepackage{stfloats}
\usepackage{url}
\usepackage{verbatim}
\usepackage{graphicx}
\usepackage{cite}

\begin{document}

\title{Observation of stable components of the sound field in Lake Kinneret using the autoproduct transform}
\author{A.L. Virovlyansky\\
\textit{{\small A.V. Gaponov-Grekhov Institute of Applied Physics of the Russian Academy of Sciences}}\\
\textit{{\small 46 Ul'yanov Street, 603950 Nizhny Novgorod,
			Russia}} \\
\textit{{\small virovlyansky@mail.ru}}\\
\date{}} \maketitle

\maketitle

\begin{abstract}
An analysis was conducted of broadband sound pulses received by a vertical array in Lake Kinneret (Israel). For most frequencies within the pulse frequency bands, the array is sparse. The application of the autoproduct transform made it possible to approximately reconstruct the signals that would be received after the emission of pulses at low frequencies for which the array is dense. Using the coherent state method developed in quantum theory, a transition has been made from representing the reconstructed field as a function of depth and time to its distribution in the 'depth-angle-time' phase space.  Due to the absence of multipath, the intensity distribution in this space should be weakly sensitive to variations in environmental parameters.  In accordance with this expectation, the distribution found is close to the result of its calculation using an idealized (range-independent) waveguide model.  It has been shown that this intensity distribution can be used as input data for a neural network when solving the problem of sound source localization in an underwater waveguide. In the examples considered, the neural network is trained on synthetic data, i.e., data obtained from theoretical calculations.
\end{abstract}

\section{Introduction}

A major factor that significantly complicates the calculation of sound fields in underwater waveguides is multipath propagation \cite{BL2003,Etter2018}. It results in the field intensity within the waveguide being described by a function with numerous local extrema, which is sensitive to minor variations in the environmental parameters. Given the inevitable inaccuracies in the mathematical model of the medium, the primary task becomes the search for and analysis of those characteristics of wave fields which, even under conditions of multipath propagation, remain relatively robust to perturbations in the waveguide parameters.

Well-known examples of such characteristics are the travel times of sound pulses coming via the individual ray trajectories \cite{BL2003,MW79,SCIENCE98}, as well as interference patterns, or striations, often seen in underwater acoustic spectrograms from broadband sources \cite{BL2003,Chuprov82,JKPS2011,DSpain99,Harrison2011}.

In works \cite{V2017,V2020a}, it was shown that the field components formed by narrow beams of rays are stable against the influence of sound speed fluctuations. The contributions of individual beams to the total field at the aperture of a receiving vertical array can be isolated using an acoustic analogue of the coherent state expansion developed in quantum theory \cite{Glauber2007,Klauder,Shl2001}. The application of this expansion allows one to move from the traditional description of the intensity of the transient sound field on a vertical receiving aray as a function of depth and time to the distribution of intensity in the phase space 'depth -- angle -- time'. The field intensity at each point of this space is formed by a narrow beam of rays. It is fundamentally important that beams whose central rays arrive at the same depth form field intensities at different points in phase space,
and the coherent state expansion resolves (at least partially) their contributions. This weakens the influence of multipath propagation, and therefore the intensity distribution in phase space is less sensitive to variations in medium parameters than the distribution in the 'depth -- time' plane.

The possibility of adequately describing the intensity distribution in phase space using an idealized environmental model is demonstrated in Ref. \cite{V2023}. This paper presents the results of processing field measurements taken at Lake Kinneret in 2019 and 2021. A monopole source emitted broadband chirp pulses, which were recorded by a vertical receiving array of 10 hydrophones spaced 3 m apart. For most of the emitted frequencies, the array was sparse. Therefore, only low-frequency components of the received signals were used in data processing. Although the sparseness of the array introduced an error, it was shown that the theoretical and experimental intensity distributions are close, and this fact can be used in solving the source localization problem.

This paper presents an analysis of the same data as in
\cite{V2023}, performed with the same purpose of comparing the calculated and measured intensity distributions in phase space.  However, an additional step is introduced here during data processing: raw signals recorded by array elements undergo a transformation called autoproduct \cite{Dowling2017,Dowling2018,Dowling2022}. Applying this transformation allows one to approximately reconstruct signals that would be received after emitting pulses at such low frequencies that the array is dense for them. The band of these reconstructed pulses starts at 0 Hz and includes frequencies that are absent in the spectrum of the actually emitted signals. The possibility of such a reconstruction is explained by the independence of the ray eikonals and amplitudes from frequency \cite{Dowling2015}. This issue is discussed further in the main part of the paper. The use of the autoproduct transformation improved the agreement between theory and experiment. 

Examples demonstrate the feasibility of using phase-space intensity distributions as input to a neural network (NN) used to solve the problem of sound source localization in a waveguide. The ability to theoretically predict intensity distributions using an existing environment model can be exploited when training NNs on synthetic data.

The paper is organized as follows. Section \ref{sec:experiment}
describes the experiments and the data obtained. The main provisions of the
theory used in the construction of the phase space representation
are outlined in Sec. \ref{sec:theory}. Section \ref{sec:autoproduct} describes the autoproduct transformation. The results of data processing performed using this transformation are presented in Sec. \ref{sec:vs}. Here, the calculated and measured distributions of the field intensity in the phase space 'depth-angle-time' are compared. Section \ref{sec:localization} examines examples of using the intensity distribution studied in this work as input to NN in solving a source localization problem. The results of the work are summarized in Sec. \ref{sec:Conclusion}.

\section{Experimental data and environmental model \label{sec:experiment}}

This section briefly repeats the description of experiments in Lake Kinneret (the Sea of Galilee) given in \cite{V2023}, the data from which are analyzed below. It also describes the mathematical models of the environment used for numerical simulation of the recorded sound fields.

Acoustic measurements in 2019 and 2021 were carried out in the central part of the lake where the bottom is almost flat. The signals were recorded using a receiving vertical array of 10 hydrophones, and the radiation was performed by a source lowered from a drifting vessel. The Table I shows the distances from which the signals were recorded.

\begin{table}
\begin{center}
\caption{Observation distances}%
\begin{tabular}
[c]{|c||c|}\hline
\text{Year} & \text{Distances}\\\hline
\text{2019} & \text{340 m, 380 m}\\\hline
\text{2021} & \text{415 m, 905 m, 1445 m}\\\hline
\end{tabular}
\end{center}
\end{table}

In 2019, at each of the indicated distances, a source located at a depth of 10 m emitted chirp pulses with a duration of 1 s in the frequency band from 300 to 3500 Hz. The receiving hydrophones covered the depth interval from 10 to 37 m with a step of 3 m. The sound speed profile measured near the receiving array is shown on the left side of Fig. 1. The dots show the depths of the receiving hydrophones.

In 2021, at the distances indicated in the table, the source was located at a depth of 7 m and emitted chirp pulses with a duration of 5 s in the frequency band from 200 Hz to 10 kHz. The sound speed profile and hydrophone depths are shown on the right side of Fig. 1. The hydrophones covered the depth interval from 7 to 34 m with a step of 3 m.

\begin{figure}[!t]
	\centering
	\includegraphics[width=4.5in]{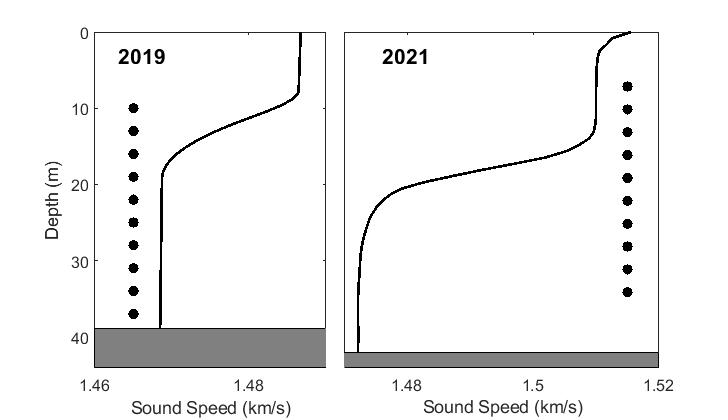}
	\caption{Sound speed profiles and hydrophone positions in 2019
		(left panel) and 2021 (right panel) (after \cite{V2023}).} \label{fig_1}
\end{figure}

When simulating the measured acoustic fields, we used range-independent
waveguide models with sound speed profiles from the left and right sides of
Fig. 1. Due to the difference in water surface levels, the waveguide depths in
2019 and 2021 were slightly different (this is shown in Fig. 1). In
simulation, they were taken equal to 38.9 m and 42.1 m, respectively.

The bottom structure of Lake Kinneret is complex \cite{Katsnelson}. For our purpose in this paper, we can use a simplified geoacoustic model that represents a liquid homogeneous half-space. Our objective here is to analyze the field components formed by narrow beams of rays whose grazing angles near the bottom do not exceed $42^{\circ}$. In Lake Kinneret, because of the high concentration of gas bubbles in the sediments, waves with such grazing angles are almost completely reflected from the bottom. In both of our waveguide models, the bottom is represented by a liquid half-space with a sound speed of 2 km/s and a density of 1400 kg/m$^{3}$. Although these environmental models are obviously inexact, they correctly reflect the fact that waves with the indicated grazing angles are completely reflected from the lower boundary of the waveguide. In this case, the reflection coefficients for all rays of a narrow beam are approximately equal to $V=\exp(i\phi)$, where $\phi$ is the angle whose value
cannot be predicted within the framework of our model. We assume that despite the inaccuracy of the bottom model, the desired field components can be approximately calculated up to unknown phase factors $V^{N}$, where $N$ is the number of beam reflections from the bottom. Below we will see that this is sufficient for the evaluation of the sound field intensity distribution in the phase space.

The sound field of a point source in a range-independent waveguide is a
function of distance $r$, depth $z$, and time $t$. The $z$ axis is vertically downward and the water surface is at the horizon $z=0$. At the
observation distance, that is, for a fixed $r$, we represent the complex field amplitude as the Fourier integral
\begin{equation}
v\left(  z,t\right)  =\int df~u\left(  z,f\right)  e^{-2\pi ift}.\label{v-u-0}%
\end{equation}
The calculation of Fourier components $u\left( z,f\right) $ at frequencies $f$ in the band of the emitted signal was carried out using the normal mode method \cite{BL2003,JKPS2011} using the KRAKEN program \cite{Porter_KRAKEN}.

\begin{figure}[!t]
	\centering
	\includegraphics[width=4.5in]{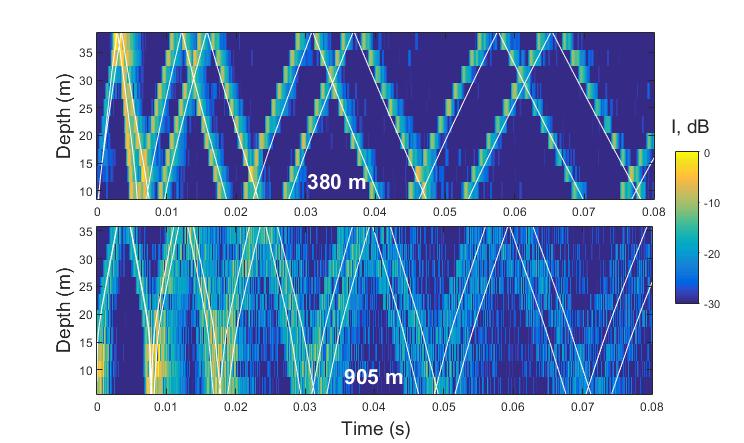}
	\caption{Signal amplitudes $|v(z, t)|$ on receiving array elements recorded from distances of 380 m (upper panel) and 905 m (lower panel). The white broken lines represent the time fronts.}
\end{figure}

When processing the measurement data, each received pulse was compressed by correlation with a replica of the transmitted signal. The functions $u\left(
z,f\right) $ and $v\left( z,t\right) $ in (\ref{v-u-0}) represent the Fourier transform and the complex amplitude of the \textbf{compressed} pulse. Figure 2 shows the time dependence of the intensities of compressed pulses $I\left( z,t\right) =\left\vert v\left( z,t\right) \right\vert ^{2}$ on the elements of the receiving array. The top and bottom panels show signals received from distances of 380 m and 905 m, respectively. In each panel, the intensity is normalized to its maximum value. Here and in what follows, for brevity, we indicate only the distance to the source, omitting the year of measurements (see Table I). In each plot, the white broken lines represent the timefront depicting the arrivals of rays in the $\left(  t,z\right)  $ plane. The timefronts were calculated
using the corresponding waveguide models. The figures show the initial
sections of the recorded signals with a duration of 0.08 s. (To avoid some confusion, we note that in Ref. \cite{V2023} similar graphs in Figs. 2 and 3 were constructed for signals filtered in the band from 300 to 900 Hz.)

\section{Theory \label{sec:theory}}

In this section, we introduce the phase space 'depth -- angle -- time' and
present relations defining the distributions of the field amplitude and
intensity in this space at a given observation distance. Here we consider the
sound field excited by a point source in a range-independent waveguide with a
sound speed profile $c\left(  z\right)  $ and a refractive index $n\left(
z\right)  =c_{0}/c\left(  z\right)  $, where $c_{0}$ is the reference sound speed.

The phase space appears in the Hamiltonian formulation of classical mechanics
and geometrical optics \cite{Gold2000,Alonso2010,Vbook2010}. Within the
framework of this formalism, the ray trajectory at distance $r$ is determined
by its depth $z$ and momentum $p=n\left(  z\right)  \sin\chi$, where $\chi$ is
the grazing angle at the point $\left(  r,z\right)  $.

In the case of a point source, all rays at $\ r=0$ escape from the same depth
$z_{s}$ with different launch angles $\chi_{0}$ and, accordingly, with
different starting momenta $p_{0}=n\left(  z_{s}\right)  \sin\chi_{0}$. The
arrival of a ray at a given observation distance $r>0$ is represented by a
point in the phase space $\left(  z,p,t\right)  $. The set of such points
forms a curve, which we call the \textbf{ray line } in the 3D phase space
$\left(  z,p,t\right)  $.

Choosing the reference sound speed $c_{0}$ = 1.5 km/s in waveguides with
$c\left(  z\right)  $ profiles shown in Fig. 1, we get $n$ values close to
one. For flat rays, the momentum $p$ is approximately equal to the grazing
angle $\chi$. Therefore, the phase space $\left(  z,p,t\right)  $ can be
called the 'depth -- angle -- time' space.

Fig. 3 shows fragments of ray lines at distances of 380 m (upper panel) and
905 m (lower panel). Both graphs are based on calculations of fans of rays
with launch angles in the range of $\pm40^{\circ}$. For a distance of 380 m, a
fan of 5000 rays is calculated, and for a distance of 905 m, a fan of 15000
rays. Each graph represents a set of points representing fan ray arrivals in
the corresponding phase space. In a refractive waveguide, fan ray arrivals
form a continuous ray line \cite{V2017,V2020a,V2020}. In our example,
discontinuities arise due to the fact that when reflected from the boundary,
the ray grazing angle, and hence its momentum, changes sign.

Note that the white lines representing the time fronts in the upper and lower
panels of Fig. 2 are projections onto the $\left(  t,z\right)  $ plane of the
ray lines shown in the upper and lower panels of Fig. 3, respectively.

\begin{figure}[!t]
	\centering
	\includegraphics[width=4.5in]{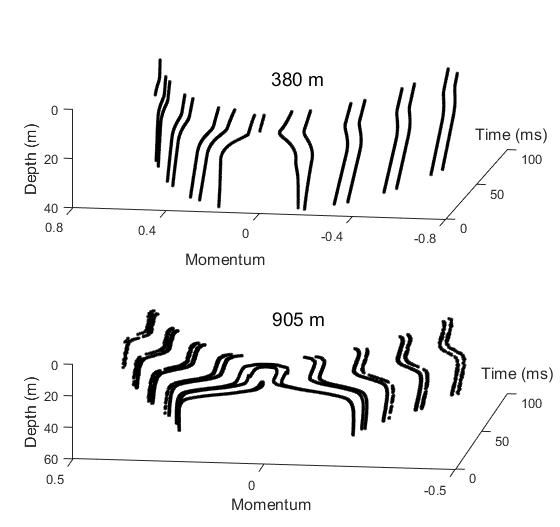}
	\caption{Fragments of ray lines at distances of 380 m (upper panel) and 905 m (lower panel) corresponding to early arrival times.} \label{fig_3}
\end{figure}

The function $u\left(  z,f\right)  $ appearing in (\ref{v-u-0}), represents
the depth dependence of the complex field amplitude at a given frequency $f$.
All our subsequent analysis is based on the transition from $u\left(
z,f\right)  $ to a function characterizing the distribution of the field
amplitude in the phase plane 'depth -- momentum (angle)' $\left(  z,p\right)
$. The transition is carried out using the coherent state expansion developed
in quantim theory \cite{Klauder,Shl2001}.

The acoustic analogue of the coherent state associated with the point $\left(
z,p\right)  $ of the phase plane is determined by the function
\begin{equation}
Y\left(  z^{\prime},f,z,p\right)  =\frac{1}{\sqrt{\Delta_{z}}}\exp\left[
ikp\left(  z^{\prime}-z\right)  -\frac{\pi\left(  z^{\prime}-z\right)  ^{2}%
}{2\Delta_{z}^{2}}\right]  ,\label{Y-mu}%
\end{equation}
where $\Delta_{z}$ is the vertical scale, $k=2\pi f/c_{0}$ is the reference
wavenumber. In quantum mechanics, a similar function describes a state with
the minimum uncertainty, that is, with the minimum product of the standard
deviations of the position and momentum \cite{LLquant}. In acoustics
(\ref{Y-mu}) describes the vertical section of a wave beam with the smallest
possible product of the beam width by the spread of the grazing angles of the
waves forming it.

When describing the field in a vertical section of a waveguide at a given
observation distance, as in \cite{V2023}, we move from $u\left(  z,f\right)  $
to the function
\begin{equation}
a\left(  p,z,f\right)  =\int dz^{\prime}~u\left(  z^{\prime},f\right)
Y^{\ast}\left(  z^{\prime},f,z,p\right)  ,\label{a-Y}%
\end{equation}
$\lambda=2\pi/k$ is the wavelength, superscript * means complex conjugate.
The complex amplitude $a\left(  p,z,f\right)  $ -- projection of $u\left(
z^{\prime},f\right)  $ onto the acoustic analogue of the coherent state
(\ref{Y-mu}) -- is formed by the contribution of a beam of rays arriving in
the depth interval $z\pm\Delta_{z}/2$ with momenta from the interval
$p\pm\Delta_{p}/2$, where $\Delta_{p}=\lambda/\left(  2\Delta_{z}\right)  $
\cite{V2020a,V2023a}. In the limiting case of high frequencies, by letting
both scales $\Delta_{z}$ and $\Delta_{p}$ tend to zero in proportion to
$\lambda^{1/2}$, one can interpret $a\left(  p,z,f\right)  $ as the
contribution to the total field of a ray arriving at a depth $z$ at a grazing
angle $\chi=\arcsin p$. Function $\left\vert a\left(  p,z,f\right)
\right\vert ^{2}$, whose quantum mechanical analogue is the Husimi function
\cite{Shl2001}, can be interpreted as the distribution of field intensity in
the phase plane $\left(  z,p\right)  $.

The fundamental property of Hamiltonian dynamics that is key for us is that
there is no multipath propagation in phase space \cite{Gold2000}. It follows
from this that no more than one ray hits any point in the phase plane. At a
finite wavelength, contributions from not individual rays but rather from
beams of rays entering phase-plane cells of area $\Delta_{z}\Delta_{p}%
=\lambda/2$ are resolved. For small $\lambda$, by optimally choosing the scale
$\Delta_{z}$ \cite{V2020a}, it is possible to achieve relatively high
resolution, allowing, although not completely, but still separating the
contributions of beams of rays coming from different directions. Such a
reduction of multipath effects makes the intensity distribution in the phase
plane $\left\vert a\left(  p,z,f\right)  \right\vert ^{2}$ significantly more
robust to perturbation than the depth dependence of the intensity $\left\vert
u\left(  z,f\right)  \right\vert ^{2}$ \cite{V2023a}.

Let us introduce the function
\begin{equation}
b\left(  z,p,t\right)  =\int df~a\left(  p,z,f\right)  e^{-2\pi ift},
\label{a-pzt}%
\end{equation}
characterizing the distribution of the transient field amplitude in phase
space 'depth -- angle (momentum) -- time' $\left(  z,p,t\right)  $. For points
of this space lying on a ray line, the function $b\left(  z,p,t\right)  $ can
be interpreted as the complex amplitude of a sound pulse that comes to depth
$z$ at grazing angle $\arcsin p$ at time $t$.

In what follows, when processing the experimental data, the main attention
will be paid to the analysis of the coherent state intensity
\begin{equation}
J\left(  z,p,t\right)  =\left\vert b\left(  z,p,t\right)  \right\vert
^{2}.\label{J-def}%
\end{equation}
At sufficiently high frequencies, this function takes its largest values near
the ray line and decreases as it moves away from it. Thus, the intensity
distribution $J$ is localized in the vicinity of the ray line and represents
its "fuzzy" version. In the phase space $\left(  z,p,t\right)  $, as in the
phase plane $\left(  z,p\right)  $ discussed in the previous section, there is
no multipath. When working with short pulses (we are talking about compressed
pulses), the resolution of the contributions of ray beams arriving from
different directions is ensured not only by spatial filtering (\ref{a-Y}) but
also by additional selection based on arrival times.

\section{Autoproduct \label{sec:autoproduct}}

Section \ref{sec:experiment} states that the inter-element distance in the receiving array used in both experiments was 3 m. To perform the transformation (\ref{a-Y}) of the recorded fields, the inter-element distance should not significantly exceed half the wavelength. Therefore, although pulses with a bandwidth of 300 - 3500 Hz were emitted in 2019 and 200 - 10000 Hz in 2021, in \cite{V2023} we limited ourselves to analyzing the components of the received signals at frequencies not exceeding 900 Hz.

In this paper, we use a different approach based on a remarkable method for transforming recorded signals called autoproduct \cite{Dowling2017,Dowling2018,Dowling2022,Dowling2015}. The version of this transform used here is expressed by the relation%

\begin{equation}
U\left(  z,f\right)  =\int_{F_{1}}^{F_{2}}dF\;u\left(  z,F+f\right)  u^{\ast
}\left(  z,F\right)  , \label{autoproduct}%
\end{equation}
where $0<F_{1}<F_{2}$ and it is assumed that the frequency interval $\left[ F_{1},F_{2}+f\right] $ is included in the band of the emitted signal. The function $U\left( z,f\right) $ is expected to mimic the field component $u\left( z,f\right) $. The result of calculating $U\left( z,f\right) $ is important for us not only at frequencies $f$ in the interval $[f_{\min},f_{\max}]$, where $f_{\min}$ and $f_{\max}$ are the lower and upper boundaries of the radiation band, but also at frequencies in the interval $[0,f_{\min}]$. Let's dwell on this in more detail \cite{Dowling2017,Dowling2015}.

The field recorded in each of the experiments under consideration is modeled as the field of a point source emitting a chirp pulse, the spectrum of which we denote by $s\left( f\right) $. The spectrum of the autocorrelation function of this pulse is equal to $\left\vert s\left( f\right) \right\vert ^{2}$. For a chirp pulse, we can approximately assume that within the frequency band $[f_{\min},f_{\max}]$ the function $\left\vert s\left( f\right) \right\vert ^{2}$ takes a constant value $S$, and outside this band it vanishes. In this approximation, the expression for the function $u\left( z,f\right) $ in the geometric optics approximation has the form

\begin{equation}
u\left(  z,f\right)  =S\sum_{\nu}A_{\nu}\left(  z\right)  e^{2\pi ift_{\nu}\left(
z\right)  }, \label{u-geom}%
\end{equation}
 where the index $\nu$ numbers the eigenrays that arrive at a depth of $z$ at the observation distance, $A_{\nu}$ and $t_{\nu}$ are the frequency-independent amplitude and arrival time of the $\nu$th eigenray. Substituting (\ref{u-geom}) into (\ref{autoproduct}) and omitting the argument $z$ of $A_{\nu}$ and $t_{\nu}$ for brevity, we find
\begin{equation}
U\left(  z,f\right)  =\left(  F_{2}-F_{1}\right)  S\sum_{\nu}\left\vert
A_{\nu}\right\vert ^{2}e^{2\pi ift_{\nu}}+G, \label{U-auto}%
\end{equation}
where
\[
G=S\sum_{\nu_{1}\neq \nu_{2}}A_{\nu_{1}}A_{\nu_{2}}^{\ast}\int_{F_{1}}^{F_{2}%
}dF~e^{2\pi if\left(  t_{\nu_{1}}-t_{\nu_{2}}\right)  }~
\]%
\begin{equation}
=S\sum_{\nu_{1}\neq \nu_{2}}A_{\nu_{1}}A_{\nu_{2}}^{\ast}e^{\pi i\left(  F_{2}%
+F_{1}\right)  \left(  t_{\nu_{1}}-t_{\nu_{2}}\right)  }\frac{\sin\left[
\pi\left(  t_{\nu_{1}}-t_{\nu_{2}}\right)  \left(  F_{2}-F_{1}\right)  \right]
}{\pi\left(  t_{\nu_{1}}-t_{\nu_{2}}\right)  }. \label{G}%
\end{equation}
If the width of the frequency integration interval, $F_2-F_1$, is sufficiently large, the term $G$ on the right-hand side of (\ref{U-auto}) can be neglected. Comparing the resulting expression with (\ref{u-geom}), we see that $u\left( z,f\right)$ and $U\left( z,f\right)$ are formed by sums of the same number of terms with different amplitudes but the same eikonals.

Since the spatial-angular structure of the field component formed by a narrow beam of rays is determined mainly by the ray eikonals, we can expect that the intensity distribution $J\left( z,p,t\right)$ of interest to us, calculated using Eqs. (\ref{a-Y}), (\ref{a-pzt}) and (\ref{J-def}) with $u\left( z,f\right)$ in (\ref{a-Y}) replaced by $U\left( z,f\right)$, mimics the distribution corresponding to the true function $u\left( z,f\right)$.

The available measurement data in Lake Kinneret allows us to find the function $u\left(z,f\right)$ only at frequencies of $f > 300 $ Hz in 2019 and $f > 200$Hz in 2021. Using the transformation (\ref{autoproduct}), we get the opportunity to reconstruct the function $u\left(z,f\right) $ in the frequency range starting from 0 Hz. This makes it possible to calculate the intensity distribution at low frequencies, for which our antenna array is dense.

When processing the 2019 data, the frequency spectra $u\left( z,f\right) $ obtained for each array element were transformed using (\ref{autoproduct}) with $F_{1}$ = 300 Hz and $F_{2}$ = 2800 Hz. The values of $U\left( z,f\right) $ were calculated for frequencies $f$ in the range from 0 to 700 Hz. The next step was the transition from $U\left( z,f\right) $ to the function

\begin{equation}
\tilde{u}\left(  z,f\right)  =U\left(  z,f\right)  \exp\left(  -\frac
{\pi\left(  f-f_{c}\right)  ^{2}}{\Delta_{f}^{2}}\right)  , \label{ut}%
\end{equation}
where $f_{c}$ = 300 Hz and $\Delta_{f}$ = 250 Hz. Then $\tilde{u}\left( z,f\right) $ was substituted into (\ref{a-Y}) instead of $u\left( z,f\right) $. This allows us to reconstruct the intensity distribution $J\left( z,p,t\right) $ in the situation where a point source emits a sound pulse
\begin{equation}
s(t)=S\Delta_{f}\exp\left(  -2\pi if_{c}t-\pi\Delta_{f}^{2}t^{2}\right)  .
\label{s-t}%
\end{equation}

The 2021 data were processed in the same way. The only difference was the use of different integration interval boundaries in (\ref{autoproduct}): $F_{1}$ = 200 Hz and $F_{2}$ = 5000 Hz.

\section{Theory vs experiment \label{sec:vs}}

In this section, we compare the intensity distributions $J\left( z,p,t\right)$ obtained by processing the measurement data with the results of theoretical calculations. The vertical scale of the acoustic analog of the coherent state $\Delta_{z}$ is taken to be 6 m. This value, as well as the above-mentioned frequency interval boundaries, $F_{1}$ and $F_{2}$, were selected empirically. The value of $\Delta_{z}$ determines the resolution scale of the intensity distribution over depth. At the central frequency of the analyzed signals $f_{c}$, the momentum resolution is determined by the parameter $\Delta_{p}=0.42$. This parameter corresponds to the angular resolution scale $\Delta_{\chi} = \arcsin\Delta_{p} = 24^{\circ}$. Time resolution scale $\Delta_{t} = 1/\Delta_{f} = $ 40 ms.

The theoretical calculation is based on solving the Helmholtz equation, which defines the field of a point source. The equation is solved using the normal mode method and the medium models described in Section \ref{sec:experiment}. The theoretical calculation of the field intensity distribution in phase space is performed in two ways.

(i) \textbf{Direct calculation of the field on the receiving array}. The complex amplitudes of the field on the array elements at the observation distances were calculated in the frequency band 15 -- 700 Hz (there are no propagating modes at frequencies below 15 Hz). These amplitudes, multiplied by $\exp\left( -\pi\left( f-f_{c}\right) ^{2} /\Delta_{f}^{2}\right) $ (as in (\ref{ut})), are used as $u\left( z,f\right) $. The amplitudes $a\left( p,z,f\right) $, as in working with measurement data, are calculated using the discrete analog of (\ref{a-Y}): integration over $z$ is replaced by summation over the depths of the array elements. Since the calculation is performed at sufficiently low frequencies, the use of denser depth grids gives close results. The desired distribution $J(p,z,t)$ is then found using Eqs. (\ref{a-pzt}) and (\ref{J-def}).

(ii) \textbf{Calculation using the autoproduct transform.} The complex field amplitudes at the array elements in the 2019 and 2021 waveguide models are calculated in the frequency ranges of 300 to 3500 Hz and 200 to 5700 Hz, respectively. Up to a constant factor, these amplitudes at each observation distance model the functions $u(z,f)$, which are found during data processing by the Fourier transform of the compressed pulses, as described in Section \ref{sec:experiment}. The functions $u(z,f)$ obtained in this way are transformed using the same relations given in section \ref{sec:autoproduct} that are used in processing measurement data for reconstructing the low-frequency components of the recorded signals and the subsequent calculation of the distribution $J(p,z,t)$.

Figures 4-9 show sections of the distributions $J(p,z,t)$ at distances of 380 m and 905 m by planes corresponding to fixed values of depth $z$ (Figs. 4 and 7), arrival time $t$ (Figs. 5 and 8) and momentum (arrival angle) $p$ (Figs. 6 and 9). The plots present two theoretical results: those from direct wave field calculation on the array elements and those using the autoproduct transform. They are labeled "Theory direct" and "Theory autoproduct," respectively.

\begin{figure}[!t]
	\centering
	\includegraphics[width=4.5in]{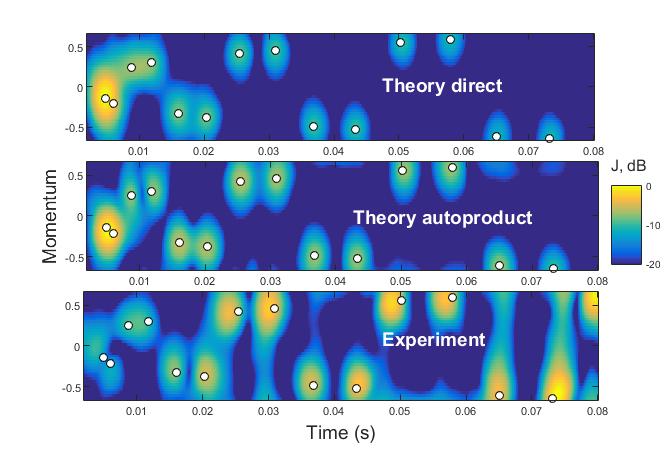}
	\caption{Cross section of the intensity distribution $J(z, p, t)$ at a distance of 380 m by the plane $z = 20$ m. White circles indicate the intersection points of the cross-section plane and the ray line. Top and middle panels: Numerical simulations. Bottom panel: Experiment.} \label{fig_4}
\end{figure}

Fig. 4 shows sections of the intensity distributions at a distance of 380 m by the z = 20 m plane, obtained by direct field calculations (upper panel), calculations using the autoproduct (middle panel), and processing of experimental data (lower panel). The similarity of the distributions in the upper and middle panels indicates that the $J(p,z,t)$ distribution obtained using the autoproduct closely mimics the true intensity distribution. Both theoretical distributions are similar to the one shown in the lower panel.

The white circles in Fig. 4 and all subsequent plots in this section indicate the intersection points of the cross-section plane and the ray line. In this example, these are the intersections with the ray line in the top panel of Fig. 3. According to the comment after (\ref{J-def}), when operating at sufficiently high frequencies, the intersection points of the ray line with any plane should be at the centers of local maxima of the cross-section. In this example, we see that this is true for most of the maxima.

Fig. 5 shows cross-sections of the same distribution at a distance of 380 m by the plane $t$ = 0.036 s. The results of theoretical calculations with and without the autoproduct transform are again similar to each other and to the distribution obtained by processing the experimental data. Here, however, the distribution maxima associated with the intersection points of the plane with the ray line are very broad and overlap. This is due to the low angular resolution, which is caused by the fact that the antenna's sparseness forces us to operate at low frequencies.

\begin{figure}[!t]
	\centering
	\includegraphics[width=4.5in]{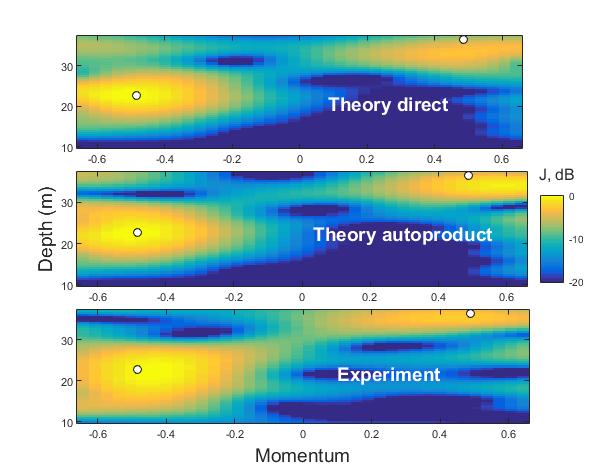}
	\caption{The same as in Fig. 4, but for a section by plane $t = 0.036$ s.} \label{fig_5}
\end{figure}

The low angular resolution is even more evident in Fig. 6, which shows a cross-section of the $J(p,z,t)$ distribution by the $p$ = 0.57 plane. Here, there are quite a few regions of relatively high intensity that are located far from the intersection points with the ray line.

\begin{figure}[!t]
	\centering
	\includegraphics[width=4.5in]{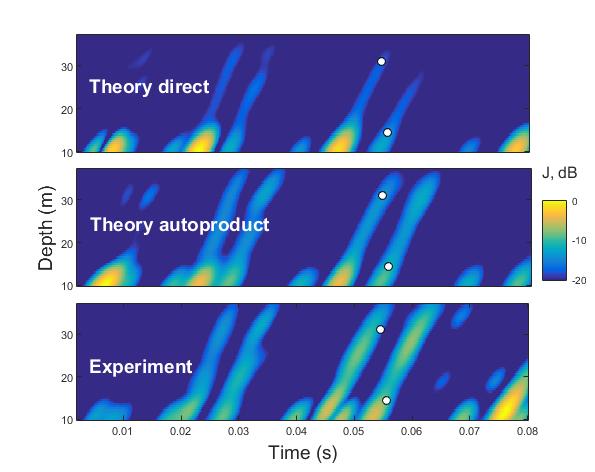}
	\caption{The same as in Fig. 4, but for a section by plane $p = 0.57$.} \label{fig_6}
\end{figure}

Figs. 7, 8, and 9 are analogs of Figs. 4, 5, and 6, respectively, for a distance of 905 m. They show sections by planes $z$ = 27 m (Fig. 7), $t$ = 0.046 s (Fig. 8), and $p$ = 0.31 (Fig. 9). In these figures, we again see that the results of the theoretical calculation with and without the autoproduct remain similar to each other and to the result obtained by processing the experimental data.

The same applies to the calculations of the cross-sections of the distributions $J(p,z,t)$ at distances of 380 and 905 m by planes corresponding to other values of $z$, $t$, and $p$, as well as similar cross-sections at our three other distances of 340, 415, and 1445 m (not shown).

\begin{figure}[!t]
	\centering
	\includegraphics[width=4.5in]{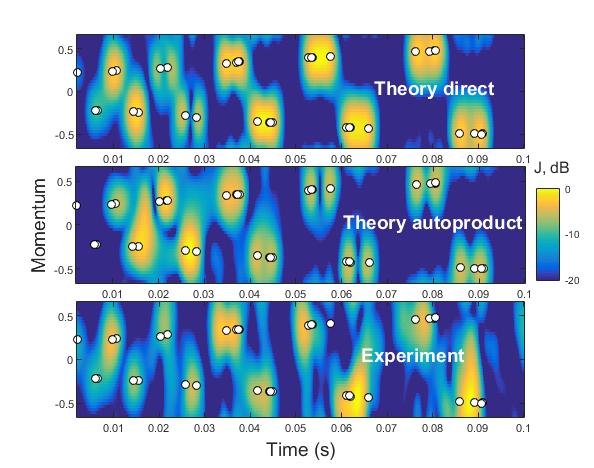}
	\caption{Cross section of the intensity distribution $J(z, p, t)$ at a distance of 905 m by the plane $z = 27$ m. White circles show the points of intersection of the plane and the ray line. Top and middle panels: Numerical simulations. Bottom panel: Experiment.} \label{fig_7}
\end{figure}

\begin{figure}[!t]
	\centering
	\includegraphics[width=4.5in]{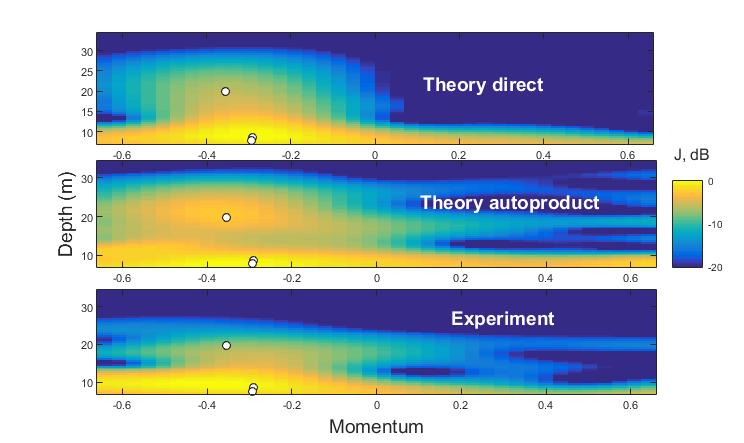}
	\caption{The same as in Fig. 7, but for a section by plane $t = 0.046 $ s.} \label{fig_8}
\end{figure}

\begin{figure}[!t]
	\centering
	\includegraphics[width=4.5in]{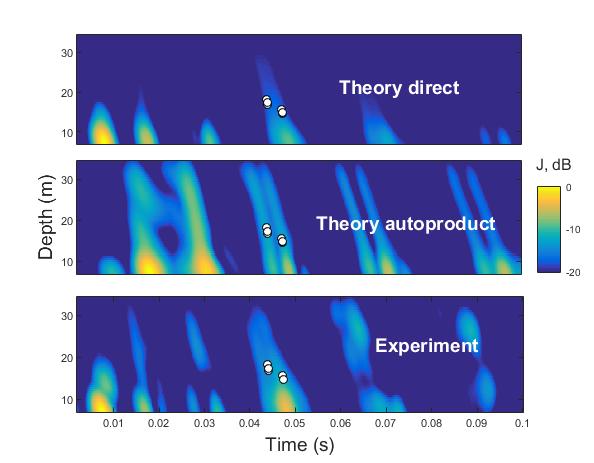}
	\caption{The same as in Fig. 7, but for a section by plane $p = 0.31$.} \label{fig_9}
\end{figure}

\section{Source localization \label{sec:localization}}

This section considers the possibility of using the $J\left( z,p,t\right)$ distribution to solve the source localization problem, that is, to estimate the source coordinates based on acoustic measurement data.

The traditional approach to solving this problem is based on matched field processing (MFP) \cite{Baggeroer, Tolstoy, Byun}. Its idea is to compare the vector of measured complex amplitudes of signals $\mathbf{u}$ at the array elements with similar vectors theoretically calculated for a set of trial source positions. The coordinates of the position corresponding to the signal vector closest to $\mathbf{u}$ are taken as the desired estimates of the source distance and depth. The normalized scalar product of the measured and calculated vectors is used as a measure of their closeness.

In the last 5-10 years, with the widespread use of machine learning methods in ocean acoustics, studies have appeared in which the localization problem is solved using a neural network (NN) \cite{Bianco,survey2023}.
The NN input is typically a tensor $\mathbf{J}$, whose elements are sampled values of some function of the field recorded by the receiving array. The elements of $\mathbf{J}$ can be, for example, the real and imaginary parts of the elements of the field correlation matrix \cite{Gerstoft2017,Niu,Wang2018} or the pressure amplitudes at the antenna elements \cite{Liu2020}. The NN essentially compares the measured tensor $\mathbf{J}$ with similar tensors -- theoretically calculated or measured -- corresponding to a set of trial source positions. As in MFP, the position of the trial point corresponding to the tensor $\mathbf{J}$ most similar to that registered by the antenna is taken as the estimate of the source position.

We have seen (Sec. \ref{sec:vs}) that the theoretical calculation of the function $J\left(z,p,t\right)$ reproduces well the field intensity distributions in phase space obtained by processing the measurement data. This is consistent with our expectation that this distribution is weakly sensitive to the inaccuracies of the medium model. In this section, we test the hypothesis that when solving the localization problem using NN, the sampled function $J\left(z,p,t\right)$ can be used as the tensor $\mathbf{J}$.

For each of our five acoustic paths, the 3D tensor $\mathbf{J}$ represents the intensity distribution in the $\sigma$ region of the phase space corresponding to the depth range $z$ covered by the receiving array (its width in both 2019 and 2021 is 27 m) and the momentum range $-\sin40^{\circ}<p<\sin40^{\circ}$. There are $N_{z}=28$ and $N_{p}=19$ sampling points, uniformly covering the specified depth and momentum intervals, respectively. The width of $\sigma$ along the $t$ axis on the 340 and 380 m long paths is 0.08 s, and on the 415, 905, and 1442 m long paths it is 0.09, 0.1, and 0.11 s, respectively. On each path, the beginning of $\sigma$ on the $t$ axis coincides with the beginning of the received compressed signal, and sampling points cover the specified intervals in 0.1 ms increments. Thus, the number of sampling points in time, $N_{t}$, varies from 801 to 1001.

For each path, sets of $M_{z}$ source trial depths $z_{s}$ and $M_{r}$ source-to-antenna trial distances $r_{s}$ are input. For each of the $M_{z}\times M_{r}$ pairs, the tensor $\mathbf{J}$ is calculated using the environment model described in Section \ref{sec:experiment}, and the NN must "decide" which of the calculated tensors is most similar to the one obtained by processing the experimental data. When solving this classification problem, we follow the procedure described in \cite{Gerstoft2017}. The differences are that instead of a correlation matrix, the NN input is fed an intensity distribution in the phase space and NN includes a convolutional layer.

\begin{figure}[!t]
	\centering
	\includegraphics[width=4.5in]{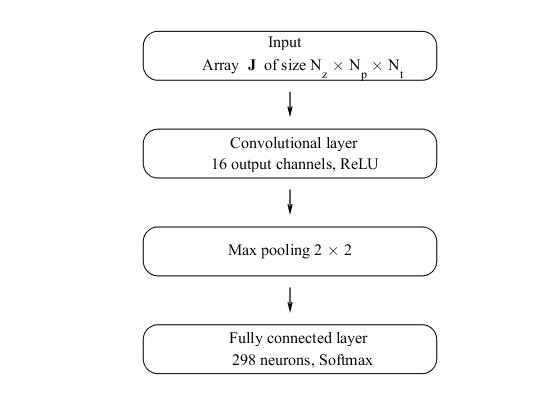}
	\caption{Neural network architecture.} \label{fig_10}
\end{figure}

The architecture of NN is shown in Fig. 10. The network is designed to solve a classification problem based on the 3D input array $\mathbf{J}$ with dimensions $N_{z}\times N_{p}\times N_{t}$. The input first passes through a convolutional layer that has a single input channel and produces 16 output channels. This convolutional layer uses a kernel size of 3 and applies padding of 1 to preserve the spatial dimensions. A ReLU activation function follows the convolution to introduce non-linearity.

After the convolutional layer, the feature maps are downsampled using a
max-pooling layer with a $2\times2$ window, which reduces the spatial size and helps to capture dominant features. Following the pooling, the network uses a fully connected layer that integrates the extracted features to make predictions.

For training, the network employs the cross-entropy loss function, which is
well-suited for classification tasks. The final output layer applies the
SoftMax function to produce normalized probabilities across the target
classes. There are $M_{r}$ times $M_{z}$ classes corresponding to all possible pairs $\left( r_{s},z_{s}\right) $. The pair to which the NN assigns the highest probability yields the desired estimates of the distance, $r_{\text{est}}$, and depth, $z_{est}$, of the source.

The same $M_{z}=14$ trial depths of 2m, 3m, ..., 15m were used for all five paths. Trial distances, spaced 10 m apart, uniformly covered the distance intervals indicated in the middle column of Table 2. The number of trial distances $M_{r}$ varied from 16 to 22.

Table II presents the results of solving the localization problem using NN. The first two columns show the actual values of the source distance $r_{\text{true}}$ and source depth $z_{\text{true}}$. As already noted, the middle column indicates the intervals of trial distances. The last two columns show the obtained estimates of distances, $r_{\text{est}}$, and depths, $z_{\text{est}}$, for each path.

\begin{table}
\begin{center}
\caption{Solutions of the localization problems}%
\begin{tabular}
[c]{lllll}%
$r_{\text{true}}$, m & $z_{\text{true}}$, m & trial ranges, m & $r_{\text{est}%
}$, m & $z_{\text{est}}$, m\\
\multicolumn{1}{c}{340} & \multicolumn{1}{c}{10} & \multicolumn{1}{c}{210 --
420} & \multicolumn{1}{c}{350} & \multicolumn{1}{c}{8}\\
\multicolumn{1}{c}{380} & \multicolumn{1}{c}{10} & \multicolumn{1}{c}{210 --
420} & \multicolumn{1}{c}{390} & \multicolumn{1}{c}{13}\\
\multicolumn{1}{c}{415} & \multicolumn{1}{c}{7} & \multicolumn{1}{c}{305 --
455} & \multicolumn{1}{c}{435} & \multicolumn{1}{c}{7}\\
\multicolumn{1}{c}{905} & \multicolumn{1}{c}{7} & \multicolumn{1}{c}{830 --
1000} & \multicolumn{1}{c}{920} & \multicolumn{1}{c}{8}\\
\multicolumn{1}{c}{1445} & \multicolumn{1}{c}{7} & \multicolumn{1}{c}{1320 --
1500} & \multicolumn{1}{c}{1390} & \multicolumn{1}{c}{6}%
\end{tabular}
\end{center}
\end{table}

\section{Conclusion \label{sec:Conclusion}}

The main result of this work is a demonstration that the autoproduct transformation makes it possible to reconstruct the intensity distribution in phase space using measurement data at frequencies for which the array is not dense.  Despite the fact that the reconstructed signals have a center frequency of 300 Hz and a band of 250 Hz, they were obtained using measurement data at higher frequencies and in much wider frequency bands.

Figs. 4-9 compare the cross sections of the measured and calculated intensity distributions. These comparisons show that theoretical predictions made using an idealized model of the medium can correctly predict the positions of intensity maxima in cross-sections. The most impressive coincidences between theory and experiment are observed in the sections $z$ = const (Figs. 4 and 7). We saw a similar coincidence in \cite{V2023}. However, due to the sparsity of the array, false maxima were observed there.

Section \ref{sec:localization} presents examples of solving a source localization problem using a NN fed with the distribution $J(p,z,t)$. The ability to relatively accurately predict the distribution $J(p,z,t)$ theoretically allows training the NN on synthetic data, that is, on data from numerical calculations of this distribution. Here, we do not compare the effectiveness of this and other methods for solving the localization problem. This issue is beyond the scope of this paper.

\section*{Acknowledgment}
The author would like to express gratitude to Professor B.G. Katsnelson and Doctor A.A. Lunkov for providing the data from the in-situ measurements in the lake Kinneret.

\end{document}